# Thriving at high hydrostatic pressure: the example of ammonoids (extinct cephalopods).

Fabio Vittorio De Blasio

Department of Geosciences, University of Oslo, Norway.

Dr. Fabio Vittorio De Blasio
University of Oslo
Department of Geosciences
Box 1047 Blindern
0316 Oslo, Norway

Phone: +47 22 02 31 22
E-mail: fvb@ngi.no

*Abstract*

Ammonoids are a group of extinct mollusks belonging to the same class of the living genus *Nautilus* (Cephalopoda). In both Nautili and ammonoids, the (usually planospiral) shell is divided into chambers separated by septa that during the lifetime were filled with gas at atmospheric pressure. The intersection of septa with the external shell generates a curve called the suture line, which in living and most fossil Nautili is fairly uncomplicated. In contrast, suture lines of ancient ammonoid were gently curved and during the evolution of the group became highly complex, in some cases so extensively frilled to be considerable as fractal curves. Numerous theories have been put forward to explain the complexity of suture ammonoid lines. Calculations presented here lend support to the hypothesis that complex suture lines aided in counteracting the effect of the external water pressure. Additionally, it is found that complex suture lines diminished shell shrinkage caused by water pressure, and thus aided improve buoyancy. Understanding the reason for complex sutures in ammonoids does not only represent an important issue in paleobiology, but is also a challenging problem in the resistance of complex mechanical structures subjected to high pressure.

**Introduction**

In the marine living mollusk *Nautilus*, a number of internal walls (called septa) divide the shell into chambers filled with gas at atmospheric pressure. This configuration is neutrally or slightly negatively buoyant so that a water jet from a conduit (called hyponome) is sufficient to propel the mollusk horizontally or vertically in the water column. The intersection of septa with the external wall, the so-called suture line, is only slightly curved in most living or fossil Nautili. In contrast, this line features greater complexity in ammonoids, the close extinct relatives of Nautili. Whereas suture lines in early ammonoid species are straight or slightly curved, their patterns became extremely complex during the evolution of the group, when highly sinuous and fractal-like features evolved [1] with fractal dimension in some cases as high as 1.6-1.7 [2]. Figure 1 shows some examples of suture lines observed in specimens of various geologic periods.

Numerous hypotheses have been put forward to explain the complexity of suture lines. The earliest theories were based on the premise that high sinuosity reinforced the shell against hydrostatic pressure perpendicular to the external shell [3] or parallel to the shell [4] preventing the mollusk from the fatal risk of an implosion when diving in deep waters. Explanations based on the external pressure were further elaborated with possible variants in more recent works [5]. Alternatively, complexity in septa has been regarded as completely unrelated to external pressure. Some researchers have suggested that complex lines increased the attachment area for muscles [6]. Others have envisaged complexity either as a consequence of the morphogenetic processes, or as a viscous fingering process [7]. Recently, the pressure hypotheses have been disputed based on the results of finite-element calculations showing that septal complexity decreases shell resistance [8], although other researchers concluded differently based on comparable modeling techniques [9,10].

In this work I present analytical calculations of shell deformation and stress with variable shell complexity. Because of the simplifications made, calculations will address the simplest, low-complexity geometries where suture lines are more or less parallel like in Manticoceras (upper drawing in figure 1). Given a certain area of the external shell comprised between two consecutive sutures (the shaded area in figure 1) the suture length will be used as a proxy for suture complexity. After showing that the increase in suture length determines a decrease in the stress, I calculate the shrinkage of the shell in response to hydrostatic pressures. It will be shown that the shrinkage is not negligible in an ammonoid with very thin shell and non complex suture.

**Model Calculations**

One single chamber is modeled as composed of two parallel and flat plates buttressed by two consecutive septa. The external pressure exerted by water on one chamber can be seen as an external load deforming inward the two flat plates (figure 2). As a preliminary step, I calculate the displacement of a rectangular plate of sides $a$ and $b$ subjected to an uniform pressure. As a boundary condition, the displacement is taken zero at the margins along the four sides of the plate. The differential equation for plate bending is [11]

$$D \nabla^4 w(x,y) = D\left[\frac{\partial^4 w(x,y)}{\partial x^4} + \frac{\partial^4 w(x,y)}{\partial y^4} + 2\frac{\partial^2 w(x,y)}{\partial x^2}\frac{\partial^2 w(x,y)}{\partial x^2}\right] = P \quad (1)$$

where $w(x,y)$ is the displacement of the plate, $x$ and $y$ are the coordinates on the surface of the plate, P is the pressure acting perpendicular to the plate and the flexural rigidity D is given by

$$D = \frac{Eh^3}{12(1-v^2)} \quad (2)$$

where $E$ is Young's modulus, $h$ is plate thickness and $v$ is Poisson's parameter. It is straightforward to find the bending for the rectangular plate with Fourier sinus expansion [11]

$$w(x,y) = \frac{a^4 b^4}{D\pi^4} \sum_{m=1}^{\infty} \sum_{n=1}^{\infty} \frac{q_{mn}}{\left[m^2 b^2 + n^2 a^2\right]^2} \sin\left(\frac{m\pi x}{a}\right)\sin\left(\frac{n\pi xy}{b}\right) \quad (3)$$

where the Fourier expansion coefficients are given as

$$q_{mn} = 4P \frac{\left[(-1)^m - 1\right]\left[(-1)^n - 1\right]}{m n \pi^2} \qquad (4)$$

so that upon integration one finds for the total deformation of the chamber

$$\Delta = \int_0^a dx \int_0^b dy\, w(x,y) = \frac{64}{\pi^8} \frac{a^5 b^5 P}{D} \sum_{m=0}^{\infty} \sum_{n=0}^{\infty} \frac{1}{(2m+1)^2 (2n+1)^2 \left[(2m+1)^2 b^2 + (2n+1)^2 a^2\right]^2} \qquad (5)$$

The maximum diagonal stresses are reached at the bottom of the plate (namely, in the inside of the phragmocone) and are given as

$$Max\{\tau_{xx}\} = \frac{6M_x}{h^2}; \quad Max\{\tau_{yy}\} = \frac{6M_y}{h^2} \qquad (6)$$

where the bending moments $M_x$, $M_y$ are

$$M_x = \int_{-h/2}^{h/2} \tau_{xx} z\, dz = -D\left[\nu \frac{\partial^2 w(x,y)}{\partial y^2} + \frac{\partial^2 w(x,y)}{\partial x^2}\right] \qquad (7a)$$

$$M_y = \int_{-h/2}^{h/2} \tau_{yy} z\, dz = -D\left[\frac{\partial^2 w(x,y)}{\partial y^2} + \nu \frac{\partial^2 w(x,y)}{\partial x^2}\right] \qquad (7b)$$

whereas shear stresses are comparatively negligible.

It can be conjectured that if the two consecutive septa are approximately parallel as in Manticoceras (see figure 1) and not too frilled like in the most complex ammonoids, the above equation for the rectangular plate can be applied also for a curved plate. In this case, the length $a$ has to be replaced by the length of the suture line $\Gamma$, whilst b is replaced by $ab/\Gamma = A/\Gamma$, so that the area of the plate $A$ remains constant. Calculations have been reported in figure 3, where the two graphs show the maximum stress in the plate (left) and the total deformation $\Delta$ of the plate (right) as a function of the ratio $R = a/b$. The five curves for each graph have been obtained with different values for plate thickness and area $A = ab$ and for a pressure of 1MPa corresponding to approximately 100 m of water depth. From the figures one can appreciate that both the maximum stress and the total deformation $\Delta$ decrease dramatically as a function of the suture length, as an increase by a factor three-four in the ratio R determines a substantial decrease of plate deformation by about two orders of magnitude. It can be concluded that an increase in the length of a suture line would have diminished both the stress in the shell and the shrinkage volume of a chamber (the total shrinkage volume of one chamber is $2\Delta$). Note the dependence of both the stress and the deformation on the area of the plate and on the thickness. So for example, a relatively large area of 10 $cm^2$ (red, blue, and cyan) determines much higher stresses and deformations than a smaller area of 6 $cm^2$ (green and yellow). However, the result strongly depends also on the plate thickness. A thin plate wall of 0.5 mm (blue and green) causes –the other parameters being the same- much higher stresses and deformations than a 1 mm thick plate (red and green) and a very thick one (cyan, 3 mm). The figure on the left reports also a typical value (131 MPa) of the maximum tensile stress that nacre is capable to sustain [4]. A shell with the properties referring to the blue curve would have imploded for a ratio $R$ less than about seven. However, more complex sutures (higher $R$) were capable of decreasing the maximum stress under the threshold value. For a sufficiently thick shell (see e.g. the example in cyan) the yield stress is never reached, and this is the strategy used by Nautili (which, indeed, do not exhibit complex suture lines).

From the graph on the right of figure 3 one can appreciate the effect of volume decrease $\Delta$ of one single plate. The volume decrease of the whole ammonoid shell subject to external pressure P would have been

$$\Delta V = 2\, N_{eff}\, \Delta(P=1MPa)\, P(MPa) \equiv \chi P \qquad (8)$$

where $N_{eff}$ is an effective number of chambers (assumed all having the same properties) and the factor two derives from the presence of two faces in every chamber. The deformation of each plate $\Delta(P=1MPa)$ is calculated above, and $\chi = \Delta V / P$ is the shrinkage coefficient. In the following, I set $N_{eff} = 10$. An ammonoid with properties as of the red curve would have been subjected to a deformation of about 5 $cm^3$ (for small R) and to a much smaller value of about 0.15 $cm^3$ for $R \approx 15$. Table 1 shows data for the two previous case of figure 3 with an external wall area of $A = 6 cm^2$ and $A = 10 cm^2$, and shell thickness t=0.5 mm corresponding to the green and blue cases respectively. The volume loss of the entire shell (third colume in table 1) decreases markedly with increasing relative suture length. The fourth colums also reports the shrinkage coefficients for these cases.

It can be suggested that the deformation of an ammonoid with thin shell, large chamber area and small complexity was not negligible, and such to decrease the fitness of the mollusk. The loss of buoyancy due to shell shrinkage is of the order $F = \Delta V g \rho_w$ where $g$ is gravity acceleration and $\rho_w$ is water density. With a value of 5 $cm^3$ this corresponds to a force of about 0.05 N. Similar to Nautili, ammonoids were able to shift vertically by pumping water through the hyponome, with a resulting thrust force of the order

$$F_{thrust} \approx \frac{1}{2}\rho U^2 S = P_{int} S \qquad (9)$$

where $U$ is the average velocity of the water jet, $P_{int}$ is the internal pressure (in the mantle cavity) and S is the area of the hyponome opening. The value measured for squid in ref. [12] is $P_{int} = 5300\ Pa$ (which corresponds to $U = 2.6 m/s$) and with $S \approx 0.1 - 0.5\ cm^2$ (a value only guessed owing to lack of conservation of ammonoid soft parts) one gets a thrust force of the order $0.053 - 0.26\ N$. Chamberlain [13] has reported a maximum thrust for living Nautili of the order 0.01 to 1.4 N according to the size of the specimen. The thrust averaged in time is strongly diminished owing to the drag force, and falls to 0.015-0.12 N. These values are of the order of the buoyancy loss that a "badly built" ammonoid (i.e., large areas exposed to pressure, thin shell and non-complex suture line) must have experienced. In figure 3B the dark shaded area corresponds to a deformation (at 100 m depth) such that the loss of buoyancy would be higher than the thrust force (assumed value of 0.05 N). Because the only known way for an ammonoid to shift vertically was through the water jet (the old analogy with the submarine according to which Nautili and ammonoids could fill and empty the camerae at will for such purpose is known today to be untenable, [10]) an ammonoid with these characteristics would have been uncapable of conteracting buoyancy loss. However, the effect of buoyancy loss was probably important also below such values, because the mollusk would have used much energy to readjust its position with the water jet. We tentatively assume that the mollusk fitness was negatively affected when the buoyancy loss was at least 10% of $T_{max}$ (light shaded region). Looking at the example in green of figure 3B, one can notice that increasing the suture length in the model ammonoid determines a gradual passage from conditions where the buoyancy loss would be superior to the thrust force to a situation where it reduced to a negligible amount.

What might have been the consequences of buoyancy loss for an ammonoid? It can be assumed that the ammonoid shell is best suited for a certain depth $\bar{y}$ below sea level, at which level the mollusk would neither rise nor sink. Because of shell shrinkage, an ammonoid floating on a level above $\bar{y}$ would have experienced an increase in the chambers volume, causing it to rise even more, whereas an ammonoid below the optimum level would have sunk. This makes the depth $\bar{y}$ a level of unstable equilibrium. The excess deformation is proportional to the pressure difference between the equilibrium depth $\bar{y}$ and the actual depth of the ammonoid y (eq. 8) Neglecting frictional drag between the shell and water, the equation of motion giving the level $y$ of the ammonoid as a function of time can be written as

$$M \frac{d^2 y}{dt^2} = \chi (P - \bar{P}) \rho g = \chi \rho^2 g^2 (y - \bar{y}) \qquad (10)$$

which gives a solution

$$y(t) = (y_0 - \bar{y}) e^{t/\tau} + \bar{y} \qquad (11)$$

where $y_0$ is the initial depth of the ammonoid below the equilibrium depth ($y_0 > \bar{y}$, with $y$ positive downwards), and

$$\tau = \frac{1}{\rho g} \left[ \frac{M}{\chi} \right]^{1/2} \qquad (12)$$

is the characteristic time of sinking. The numerical value for $\tau$ can be estimated based on the previous calculations and on assumed masses for ammonoids.

Looking back at table 1, the numerical examples based upon the previous case of figure 3 (with an external wall area of $A = 6 cm^2$ and $A = 10 cm^2$) one can see that the characteristic time $\tau$ of a model ammonoid with these characteristics is very short for a straight line (for such rapid times the corresponding velocities becomes such that the drag force is not negligible). With a suture line of relative length 15-20, the time becomes of some minutes, showing that it would have been much more easy in this case to compensate for buoyancy loss by jet propulsion. The table 1 also shows (sixth column) that for a suture line of 6 cm in length, the force at a depth of 100 m below the optimum level is about 0.04 N, while it becomes about 0.0008 N for a suture line of length 18 cm. This force is comparable to the force arising from buoyancy loss at 100 m deviation from the optimum level for the model ammonoid (see table 1) for the case with short suture length. Instead, an ammonoid with longer suture length would have been subject to a much smaller buoyancy loss. Although this calculation is affected by major uncertainties, it points to the basic problem that an ammonoid with straight suture lines would have needed to correct its level frequently with a powerful water jet. Not only would this behavior have caused a substantial energy loss, but it is also at variance with the observations of living Nautili, which are slow swimmers and use jet propulsion only occasionally to escape predators. Additionally, the ammonoid might have reached a fatal non-return point if buoyancy loss was greater than the maximum thrust force attainable.

Note also that because the external shells of ammonoids were usually rather thin compared to modern Nautili (typically a fraction of mm compared to several mm), and because both stress and buoyancy loss increase markedly with decreasing thickness ($\Delta \propto h^{-3}$), high stress and shell shrinkage must have been a more severe problem for ammonoids than it is for living Nautili. Increasing the thickness of the shell was probably an evolutionary poor strategy, as it required a greater effort for the production of nacreous shell, which would have also become more massive. Rather, ammonoids chose the evolutionary approach of augmenting the support of the external shell, by surrounding the plate with a longer perimeter.

**Conclusions**

Analytical calculations illustrate the fundamental role of suture length in diminishing the stress and of deformation in ammonoid external shells, and are confirmed by by.finite element computations [14]. Whereas the effect of stress decrease with increasing suture length confirms previous investigations [5], shell shrinkage with consequent loss of buoyancy is a newly suggested mechanism. Complex suture lines reduced shell shrinkage caused by hydrostatic pressure, with the effect of improving buoyancy and dynamical stability of ammonoids through the water column, also at pressures lower than the point of catastrophic implosion. Note that deformation of nacreous plates of the order of the one calculated here is commonly observed in the laboratory [15]. The calculations presented here are applicable only to suture lines that are only slightly curved, and this limits our considerations to the simplest suture lines of the most primitive ammonoids like in Manticoceras (fig. 1). Thus, the more striking case of fractal suture lines cannot be addressed by the present approach. However, since the evolutionary biology of ammonoids demonstrates a certain continuity in complexity increase during geological time, it seems likely that an explanation valid for the simplest suture lines may hold also for the complex ones.

It is interesting to note that buoyancy loss resulting from pressure shrinkage is the same reason why submarine hulls are constructed with thick metal and not with low Young modulus materials such as glass fibers, which would otherwise be an attractive alternative (see for example Gordon [16] for a popular account). A final solution to the problem of ammonoid suture lines may have potential applications to the engineering of structures subjected to high pressures such as submersibles and submarines.

FIGURE CAPTIONS

Figure 1. Ammonoid suture lines. Early ammonoids had slightly sinuous suture lines (termed goniatitic) like in the Devonian genus *Manticoceras* (top). Higher levels of complexity were reached with the ceratitic suture (line in the middle of the figure, *Paraceratites,* Triassic), and then with the ammonitic suture (line at the bottom, *Perisphinctes*, late Jurassic).

Figure 2. The basic model chamber. Only the two external walls of each chambers are subjected to water pressure (as indicated by the arrow for the upper wall). The deformation of each external wall is calculated with plate theory.

Figure 3. The maximum stress in kPa experienced by one plate (left) and shrinkage volume of the plate $\Delta(P=1MPa)$ (in $cm^3$) calculated from Eq. 6 for a pressure of 1 MPa as a function of the ratio R=a/b between the length of the suture line and the distance between successive lines. The parameters for each line are the following (from the top to the bottom): blue: $h=0.5$ mm; $A=10\ cm^2$; green: $h=0.5$ mm; $A=6\ cm^2$; red: $h=1$ mm; $A=10\ cm^2$; yellow: $h=1$ mm; $A=6\ cm^2$; cyan: $h=3$ mm; $A=10\ cm^2$. The three horizontal lines in the first graph are three values for the maximum stress. From the lowest value: 29 MPa (Pfaff, ref. [4]); 131 MPa (Hewitt and Westermann, ref. [5]) and 231 MPa [15]. Note that the dependence on the shrinkage volume on the thickness is $\Delta \propto 1/h^3$. Values $E=5\cdot 10^{10} Pa$ and $\nu=0.29$ have been used in the calculations [9].

TABLE CAPTION

Table 1. Data are shown for two different plates: A: $h=0.5$ mm; $A=6\ cm^2$ (corresponding to the green color in figure 3), and B: $h=0.5$ mm; $A=10\ cm^2$ (blue color in figure 3). In both table 1A and 1B the data reported are: the dimensionless suture length R defined as the ratio of the two plate lengths $R=a/b$, $\Delta$ defined in equation 6 at a depth of 100 m below the optimum depth, the shrinkage volume $\Delta V$ for the entire shell, the shrinkage coefficient $\chi$ for the entire shell, the characteristic time $\tau$ and the buoyancy loss F. The other parameters are $N_{cham} = 10$, $M = 0.3\,Kg$.

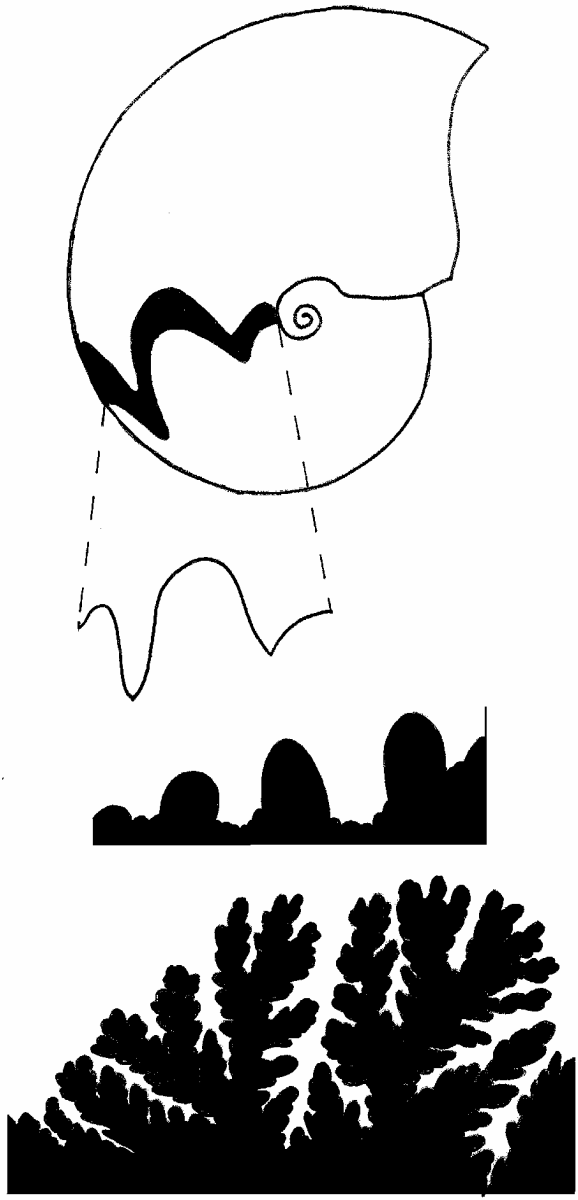

FIGURE 1

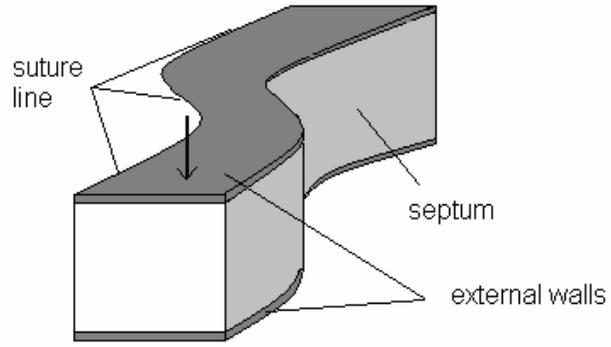

FIGURE 2

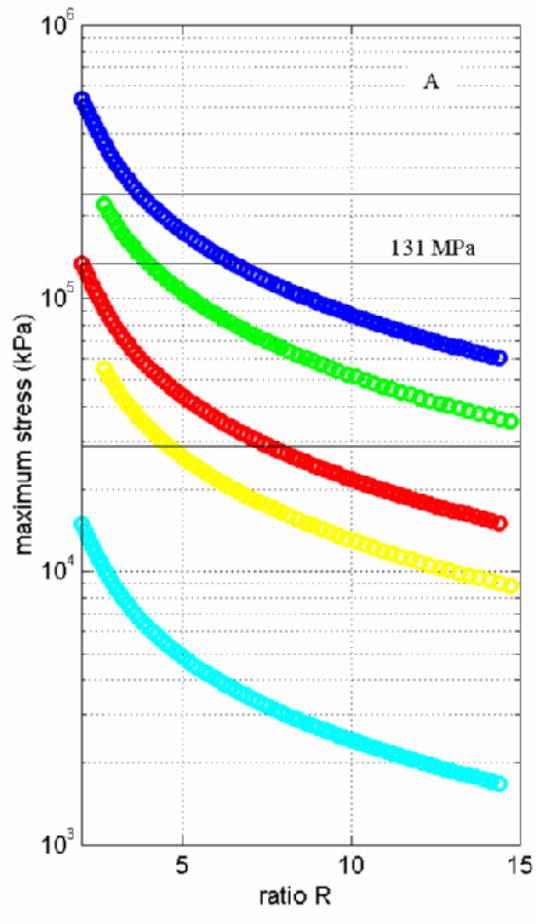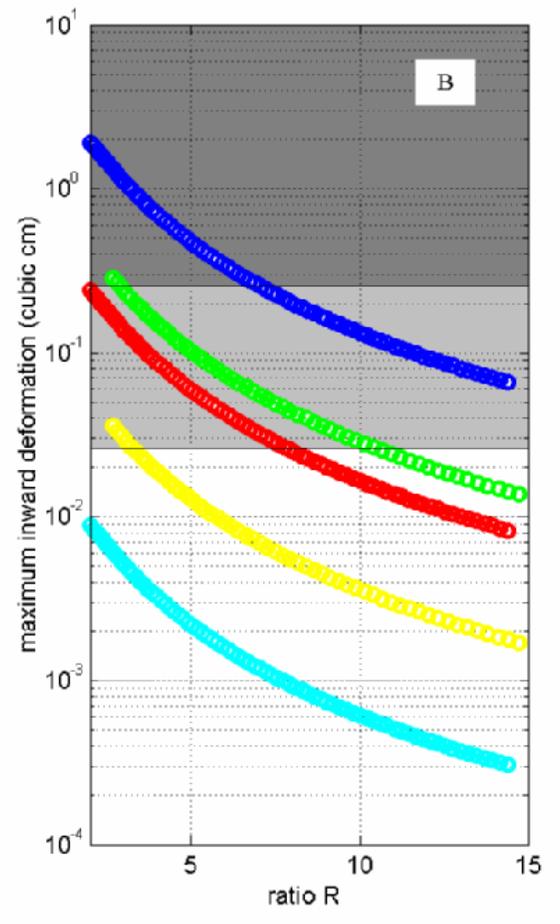

FIGURE 3

TABLE 1

| Relative length of suture R | $\Delta$ ($cm^3$) | $\Delta V$ ($cm^3$) | $\chi$ ($cm^3 MPa^{-1}$) | $\tau$ (s) $M = 0.3 Kg$ | $F$ (N) at $y - \bar{y} = 100$m |
|---|---|---|---|---|---|
| 2.66 | 0.286 | 5.72 | 5.72 | 22.86 | 0.057 |
| 10 | 0.028 | 0.56 | 0.56 | 73.1 | 0.0056 |
| 20 | 0.007 | 0.014 | 0.14 | 146.4 | 0.0014 |

A

| Relative length of suture R | $\Delta$ ($cm^3$) | $\Delta V$ ($cm^3$) | $\chi$ ($cm^3 MPa^{-1}$) | $\tau$ (s) $M = 0.3 Kg$ | $F$ (N) at $y - \bar{y} = 100$m |
|---|---|---|---|---|---|
| 2.66 | 1.373 | 27.46 | 27.46 | 10.39 | 0.274 |
| 10 | 0.132 | 2.64 | 2.64 | 33.8 | 0.026 |
| 15 | 0.061 | 1.22 | 1.22 | 49.5 | $6 \times 10^{-4}$ |

B